\documentclass{svproc}
\usepackage{url}
\usepackage{epsfig}

\begin{document}
\mainmatter              
\title{Hyperon interaction with dense nuclear matter and link to neutron stars}
\titlerunning{Hyperon interaction with dense nuclear matter and link to neutron stars}  
%
\author{Laura Tolos}
\authorrunning{Laura Tolos} 
%
\tocauthor{Laura Tolos}
\institute{Institut f\"ur Theoretische Physik, University of Frankfurt, \\ Max-von-Laue-Str. 1, 60438 Frankfurt am Main, Germany;  \\
Frankfurt Institute for Advanced Studies, University of Frankfurt, \\ Ruth-Moufang-Str. 1,
60438 Frankfurt am Main, Germany; \\
Institute of Space Sciences (ICE, CSIC), Campus UAB, \\ Carrer de Can Magrans, 08193, Barcelona, Spain; and \\
Institut d' Estudis Espacials de Catalunya (IEEC), \\ 08034 Barcelona, Spain \\
\email{e-mail: tolos@th.physik.uni-frankfurt}}
\maketitle              

\begin{abstract}
The theoretical status of the hyperon-nucleon and hyperon-hyperon interactions is reviewed, paying a special attention to chiral effective field theories. Results on hyperons in dense matter are presented and the consequences for the properties of neutron stars are analyzed.
\end{abstract}
\section{Introduction}
Over the past decades the properties of hyperons in dense matter have been object of high interest, in connection with on-going and future experiments on hypernuclei as well as the possible existence of hyperons in neutron stars. 

In order to determine the properties of hyperons in a dense medium, it is of crucial importance to have a deep understanding of the underlying bare hyperon-nucleon (YN) and hyperon-hyperon (YY) interactions, and the modifications induced in a dense medium.

In this contribution, the theoretical status of the YN and YY interactions is reviewed, concentrating the attention on the recent developments within chiral effective field theory ($\chi$EFT). The properties of hyperons in dense nuclear matter within $\chi$EFT are studied, whereas  the presence of hyperons in neutron stars and the consequences for their structure are analyzed.

\section{Theoretical approaches to YN and YY interactions} 
Despite the scarce experimental information on YN and YY interactions, there has been an important effort from the theoretical point of view to describe the YN and YY interactions. Those include meson-exchange models, $\chi$EFT approaches, calculations on the lattice, low-momentum models and quark model approaches.

{\it Meson-exchange models} rely on the fact that the interaction between two baryons is given by the exchange of mesons. In particular,
the determination of the YN and YY interactions is based on the nucleon-nucleon (NN) meson-exchange model under the assumption of $SU(3)_{\rm flavor}$ symmetry. Among those models, one should mention the J\"ulich \cite{Holzenkamp:1989tq,Haidenbauer:2005zh} and Nijmegen potentials (see \cite{Rijken:2010zzb} and references therein).

The YN and YY {\it models based on $\chi$EFT} have been also built recently by constructing a systematic approach that respects chiral symmetry. More precisely,  the J\"ulich-Bonn-Munich group has constructed the YN and YY interactions within  $\chi$EFT starting from their previous approach for the NN interaction \cite{Polinder:2006zh,Haidenbauer:2013oca,Haidenbauer:2016vfq}. In the next section we discuss this approach in more detail.

Another way of constructing the YN and YY interactions is based on solving QCD on the lattice. Within  {\it lattice QCD}, Monte Carlo methods are used to solve the QCD path integral over the quark and gluon fields at each point of a four-dimensional space-time grid. The effort in this direction has been lead by the HALQCD (see contribution to these proceedings)  and the NPLQCD \cite{nplqcd} collaborations.

Other approaches include {\it low-momentum interactions}  and {\it quark model potentials}. The former aims at obtaining a universal effective low-momentum potential for YN and YY using renormalization-group techniques \cite{Schaefer:2005fi}, whereas the latter builds the YN and YY interactions within constituent quark models \cite{Fujiwara:2006yh}.

I refer the reader to the recent review of Ref.~\cite{Tolos} and references therein.


\section{Hyperons in dense matter within $\chi$EFT}

The $\chi$EFT has been used to described NN interaction to a high precision. However, only during the past decade the YN interaction has been object of analysis within the $\chi$EFT. The YN interaction has been obtained within the SU(3) $\chi$EFT deriving the different orders in the chiral expansion and improving calculations in a systematic way by going to higher orders in the Weinberg power counting. At leading order (LO) in the power counting, the YN potential consists of single pseudoscalar-meson exchanges and non-derivative four-baryon contact terms \cite{Polinder:2006zh}, whereas the next-to-leading order (NLO) accounts for two pseudoscalar-meson exchanges and contact interactions with two derivatives \cite{Haidenbauer:2013oca,Haidenbauer:2019boi}. 

The solution of a regularized Lippmann-Schwinger equation using, as kernel, the YN LO and NLO contributions allows for the calculation of scattering observables.  In Ref.~\cite{Haidenbauer:2013oca} the NLO13 interaction was determined by fixing the baryon-baryon-meson couplings constants using the available standard YN data points and the SU(3) symmetry.  This symmetry has also helped to derive the various low-energy constants, although this symmetry is broken by using the hadron physical masses.  As a result, the available $\Lambda N$ and $\Sigma N$ data is described consistently. However, the simultaneous determination of NN and YN interactions with contact terms completely respecting SU(3) symmetry was not possible \cite{Haidenbauer:2013oca}.

The properties of $\Lambda$ and $\Sigma$ in dense matter have been later on analyzed using the LO and NLO YN interactions \cite{Haidenbauer:2014uua}. Within the Brueckner-Hartree-Fock framework, the single-particle potentials of the $\Lambda$ and $\Sigma$ hyperons in nuclear matter have been obtained. The $\Lambda$ single-particle potential has been found to be in good qualitative agreement with the empirical values extracted from hypernuclear data, whereas the $\Sigma$-nuclear potential has been determined to be repulsive. These results have been improved in the subsequent analysis of Ref.~\cite{Petschauer:2015nea}. Whereas the $\Sigma$-nuclear potential becomes moderately repulsive, the $\Lambda$ one is repulsive starting at two-to-three times saturation density.

The effect of the three-body forces has also been studied for the $\Lambda$-nuclear interaction in dense matter \cite{Haidenbauer:2016vfq}. 
The inclusion of three-body forces is needed in order to obtain, for example, the nuclear saturation in non-relativistic approaches, such as the Brueckner-Hartree-Fock.  The $\Lambda$ single-particle potential turns out to be more repulsive  when three-body forces are considered \cite{Haidenbauer:2016vfq}.

More recently, there has been a reanalysis of the work in \cite{Haidenbauer:2013oca} regarding the $\Lambda N$ and $\Sigma N$ interactions. In this recent work \cite{Haidenbauer:2019boi} the number of constants is reduced by inferring some of them from the $NN$ sector using SU(3) symmetry, leading to the NLO19 interaction. Whereas NLO13 of Ref.~\cite{Haidenbauer:2013oca} and NLO19 of Ref.~\cite{Haidenbauer:2019boi} yield equivalent results for $\Lambda N$ and $\Sigma N$ scattering observables, the in-medium $\Lambda$ and $\Sigma$ properties in matter are affected by the different choice. This is due to the fact that the strength of $\Lambda N \rightarrow \Sigma N$ transition potential changes.  The $\Lambda$ single-particle potential from the new NLO19 is much more attractive than the NLO13 one, whereas  the NLO19 interaction provides slightly more repulsion for the $\Sigma$ single-particle potential. 

With regards to $\Xi N$ interaction and the $\Xi$-hyperon in nuclear matter, the calculation within $\chi$EFT up to NLO \cite{Haidenbauer:2018gvg} shows that the $\Lambda \Lambda$ s-wave scattering length and upper bounds of the $\Xi^- p$ cross sections
are compatible with the obtained $\Xi N$ interaction.  The $\Xi$ single-particle potential in nuclear matter ranges between -3 to -5 MeV, similar to other  Brueckner-Hartree-Fock calculations, whereas the reported experimental value is larger.

\section{Hyperons in Neutron Stars}

The knowledge of the properties of hyperons in dense matter is of crucial importance in the context of neutron stars. Neutron stars are the most compact known objects without event horizons and, therefore, serve as a unique laboratory for dense matter physics \cite{Watts:2016uzu}. Their bulk features, such as mass and radius, strongly depend on the properties of matter in their interior and, hence, on the equation of state (EoS). The most precise measurements of masses are located around the Hulse-Taylor pulsar. Accurate values of approximately $2M_{\odot}$ have been determined recently \cite{Demorest:2010bx,Antoniadis:2013pzd,2019NatAs.tmp..439C}. As for radii, precise determinations do not yet exist, being the simultaneous determination of both mass and radius the focus of the ongoing NICER (Neutron star Interior Composition ExploreR) \cite{nicer} and the future eXTP (enhanced X-ray Timing and Polarimetry) missions \cite{Watts:2018iom}.

The composition of the neutron star interior is determined by demanding equilibrium against weak interaction processes, the so-called $\beta$-stability.  Traditionally the interior of neutron stars was modelled by a uniform fluid of neutron rich matter in  $\beta$-equilibrium, where the main ingredients are neutron, protons and electrons.  However, other degrees of freedom are expected, such as hyperons, because of the fast increase of the nucleon chemical potential with density as well as the fact that the density in the center of neutron stars is very high. 

The presence of hyperons affects the EoS of the neutron star interior.  The EoS and, hence, the pressure becomes softer with respect to the case when only neutrons and protons are present. The addition of one particle specie opens a set of new available low-energy states that can be filled, hence lowering the total energy of the system. Neutron stars are in hydrostatic equilibrium, that is, there exists a balance between the gravitational force and the internal pressure. Thus, the less pressure, the less mass a neutron star can sustain.

As indicated before, hyperons are energetically favoured in neutron stars. However, their presence induces a softening of the EoS. A softening of the EoS could then lead to masses for neutron stars below the $2M_{\odot}$ observations. In the literature this effect is usually referred as {\it the hyperon puzzle}. Over the years, several solutions have been put forward so as to have hyperonic $2M_{\odot}$ neutron stars. One possible solution  is based on stiff YN and YY interactions. In this manner, a stiff EoS will be produced that overcomes the softening induced by hyperons, thus leading to $2M_{\odot}$ neutron stars.  Another solution relies on the stiffening induced by hyperonic three-body forces. Nevertheless, there is not a general consensus whether models with hyperonic three-body forces will allow for $2M_{\odot}$ neutron stars.  Also other solutions have been considered, such as those assuming new degrees of freedom ($\Delta$ baryons or condensates), or the presence of a transition to quark matter below the hyperon onset with strong quark interactions to still reach $2M_{\odot}$. 

I refer the reader to the recent reviews of Ref.~\cite{Tolos,Vidana:2018bdi} and references therein.

\section*{Acknowledgements}
This research is supported by the Heisenberg Programme of the Deutsche \\ Forschungsgemeinschaft under the Project Nr. 383452331 (Heisenberg Programme) and Nr. 411563442; the Spanish Ministerio de Econom\'ia, Industria y
Competitividad under contract FPA2016-81114-P; the EU STRONG-2020 project under the program
H2020-INFRAIA-2018-1, grant agreement no. 824093; and the PHAROS COST Action CA16214.

%
%

\end{document}